\documentclass[a4paper,11pt]{article}
\usepackage{pos}

\usepackage{physics}
\usepackage{color}
\usepackage{comment}
\usepackage{ulem}

\title{Comparison with model-independent and dependent analyses for pion charge radius}

\author*[a]{Kohei Sato}
\author[b]{Hiromasa~Watanabe}
\author[c,d]{Takeshi~Yamazaki}

\affiliation[a]{Degree Programs in Pure and Applied Sciences, Graduate School of Science and Technology, University of Tsukuba, Tsukuba, Ibaraki 305-8571, Japan}
\affiliation[b]{Yukawa Institute for Theoretical Physics, Kyoto University, Kyoto 606-8502, Japan}
\affiliation[c]{Faculty of Pure and Applied Sciences, University of Tsukuba, Tsukuba, Ibaraki 305-8571, Japan}
\affiliation[d]{Center for Computational Sciences, University of Tsukuba, Tsukuba, Ibaraki 305-8577, Japan}

\emailAdd{ksatoh@het.ph.tsukuba.ac.jp}

\abstract{
Traditionally, there has been a method to extract the charge radius of a hadron based on the fits of its form factor with some model assumptions. In contrast, a completely different method has been proposed, which does not depend on the models. In this report, we explore several improvements to this model-independent method for analyzing the pion charge radius. Furthermore, we compare the results of the pion charge radius obtained from $N_{f}=2+1$ lattice QCD data at $m_{\pi}=0.51$ GeV using the three different methods: the traditional model-dependent method, the original model-independent method, and our improved model-independent method. In this comparison, we take into account systematic errors estimated in each analysis.
}

\FullConference{
The 40th International Symposium on Lattice Field Theory (Lattice 2023)
\\
July 31st - August 4th, 2023
\\
Fermi National Accelerator Laboratory
}

\begin{document}
\maketitle

\section{Introduction}
Traditional analysis of hadron form factor in lattice QCD to obtain the charge radius involves fitting the form factor assuming some models. For example, in the case of the charged pion, the electromagnetic form factor for each momentum transfer is calculated from the ratio of the 3-point function $\tilde{C}_{\pi V\pi}(t,t_{\rm{sink}};\vec{p})$ with some overall factors given by
\begin{eqnarray}
	F_{\pi}(q^2)=\frac{2E_{\pi}(\vec{p})Z_{\pi}(\vec{0})}{(E_{\pi}(\vec{p})+m_{\pi})Z_{\pi}(\vec{p})}
                    \frac{\tilde{C}_{\pi V\pi}(t,t_{\rm{sink}};\vec{p})}{
							\tilde{C}_{\pi V\pi}(t,t_{\rm{sink}};\vec{0})}e^{(E_{\pi}(\vec{p})-m_{\pi})t} ,
\end{eqnarray}
where $\vec{p}:=\frac{2\pi}{L}(n_{x}, n_{y}, n_{z})$ with $n_i$ and $L$ being an integer and the spatial extent, $E_{\pi}(\vec{p}):=\sqrt{m_{\pi}^2+\vec{p}^2}$, $Z_{\pi}(\vec{p}):=\matrixel{0}{\pi^{+}(\vec{0},0)}{E_{\pi}(\vec{p})}$. This form factor data $(q^2,F_{\pi}(q^2))$ is analyzed by fitting with some model assumptions such as monopole, polynomial, and z-expansion functions. Those models are predicted by based on specific theories for each form factor. However, a completely different method that does not rely on such models has been proposed~\cite{Aglietti:1994nx,Lellouch:1994zu,Bouchard:2016gmc,Feng:2019geu}. In this method, there is no systematic error due to fit ansatz. In the previous report~\cite{Sato:2022qee}, we found that the finite-volume effect appears for certain cases in the new approach~\cite{Feng:2019geu} and hence proposed an improvement method to mitigate it.

\par

In this report, we discuss a further improvement to our proposal, and also evaluate the pion charge radius from three methods including our proposal.
Those results are compared using systematic errors estimated in each method.

\section{Model-independent method}
The key idea of the model-independent method is 
\begin{eqnarray}
	\eval{\dv{\tilde{F}(\vec{p})}{\abs{\vec{p}}^2}}_{\abs{\vec{p}}^2=0}	
	=-\dfrac{1}{3!}\int \dd[3]{x}\abs{\vec{x}}^2F(\vec{x})
    \label{eq:moment}
\end{eqnarray}
holds for the continuum limit and infinite volume limit, where $F(\vec{x})$ satisfies $F(\vec{x})=F(\abs{\vec{x}})$ and $\tilde{F}(\vec{p})$ is the Fourier transform of $F(\vec{x})$. For example, if $F(\vec{x})$ is a 3-point function $C_{\pi V\pi}(\vec{x})$, this equation indicates that the pion charge radius can be obtained from $\abs{\vec{x}}^2$ moment of the 3-point function.

\par

Let us consider this calculation for finite volumes. To simplify the calculation, we discuss the 1-dimensional 3-point function defined by
\begin{eqnarray}
	C_{\pi V\pi}(t,t_{\rm{sink}};r):=Z_{V}\sum_{\vec{z}}\sum_{y_{2},y_{3}}\sum_{x_{2},x_{3}}
	\matrixel{0}{\pi^{+}(\vec{z},t_{\rm{sink}})V_{4}(\vec{y},t){\pi^{+}}^{\dag}(\vec{x},0)}{0}
	\label{eq:1pt3func} ,
\end{eqnarray}
where $r:=\abs{x_{1}-y_{1}}$. We assume the periodic boundary condition in all spacetime directions and the $1 \ll t \ll t_{\rm sink}$ region for the ground state dominance. Using  the Fourier transform of Eq.~(\ref{eq:1pt3func})
\begin{eqnarray}
	\tilde{C}_{\pi V\pi}(t,t_{\rm{sink}};p)
		=Z_{V}Z_{\pi}(0)Z_{\pi}(p)L^2\dfrac{(E_{\pi}(p)+m_{\pi})}{2m_{\pi}2E_{\pi}(p)}
		F_{\pi}(q^2)e^{-E_{\pi}(p)t}e^{-m_{\pi}(t_{\rm{sink}}-t)} ,
    \label{eq:1pt3func_p}
\end{eqnarray}
the normalized $n$-th moment of the 3-point function, $C^{(n)}(t):=\sum_{r}r^{2n}C_{\pi V\pi}(t,t_{\rm{sink}};r)/\tilde{C}_{\pi V\pi}(t,t_{\rm{sink}};0)$, is given by
\begin{eqnarray}
	C^{(n)}(t)=\sum_{p}\Delta(t,p)T_{n}(p)F_{\pi}(q^2)
	\label{eq:sumpDTF} ,
\end{eqnarray}
where $\Delta(t,p):=\dfrac{Z_{\pi}(p)}{Z_{\pi}(0)}\dfrac{E_{\pi}(p)+m_{\pi}}{2E_{\pi}(p)}e^{-(E_{\pi}(p)-m_{\pi})t}$, $T_{n}(p):=\dfrac{1}{L}\sum_{r}r^{2n}e^{ipr}$. From the Taylor expansion of the form factor, $F_{\pi}(q^2)=\sum_{m=0}^{\infty}f_{m}q^{2m}$, Eq.~(\ref{eq:sumpDTF}) can also be written as
\begin{eqnarray}
	\hspace{-50pt}
	C^{(n)}(t)=f_{0}\beta_{0,n}(t)+f_{1}\beta_{1,n}(t)+
		f_{2}\beta_{2,n}(t)+\cdots=\sum_{m=0}^{\infty}f_{m}\beta_{m,n}(t)
	\label{eq:cn=fbeta} .
\end{eqnarray}
The function $\beta_{m,n}(t)$ is a known function given by $\beta_{m,n}(t):=\sum_{p}\Delta(t,p)T_{n}(p)q^{2m}$. Therefore, the relation in Eq.~(\ref{eq:moment}) becomes Eq.~(\ref{eq:cn=fbeta}) for finite volume. In the infinite volume, the 1st-order spatial moment calculation exactly corresponds to the 1st-order momentum-derivative $f_{1}$, while on a finite volume, higher-order derivatives also appear other than the 1st-order derivative,  $f_{2},f_{3},\cdots$.

\par

To reduce the higher-order contamination in Ref.~\cite{Feng:2019geu}, the function $R(t)$ is defined by
\begin{eqnarray}
	R(t)&:=&\alpha_{1}C^{(1)}(t)+\alpha_{2}C^{(2)}(t)+h\notag\\
	&=&(\alpha_{1}\beta_{0,1}+\alpha_{2}\beta_{0,2}+h)
		+(\alpha_{1}\beta_{1,1}+\alpha_{2}\beta_{1,2})f_{1}
		+(\alpha_{1}\beta_{2,1}+\alpha_{2}\beta_{2,2})f_{2}+\cdots
	\label{eq:defR(t)} ,
\end{eqnarray}
where we use $f_{0}=1$ and the dots represent higher-order terms with $f_m$ ($m\ge 3$). The parameters $\alpha_{1},\alpha_{2},h$ are chosen to satisfy 
\begin{eqnarray}
	\alpha_{1}\beta_{0,1}+\alpha_{2}\beta_{0,2}+h=0,
	\hspace{15pt}\alpha_{1}\beta_{1,1}+\alpha_{2}\beta_{1,2}=1,
	\hspace{15pt}\alpha_{1}\beta_{2,1}+\alpha_{2}\beta_{2,2}=0
	\label{eq:defParaAlphaH}
\end{eqnarray}
in order to reduce the additional contribution from terms other than $f_{1}$. From Eqs.~(\ref{eq:defR(t)}) and (\ref{eq:defParaAlphaH}), the function $R(t)$ is rewritten as
\begin{eqnarray}
	R(t)=f_{1}+\sum_{m=3}^{\infty}\qty(\sum_{k=1}^{2}\alpha_{k}\beta_{m,k}(t))f_{m}.
	\label{eq:R(t)}
\end{eqnarray}
The first term in Eq.~(\ref{eq:R(t)}) is the value of the 1st-order derivative of the form factor that we want to obtain, and the second term is the time-dependent high-order contamination due to the finite volume. In the following, we call this the original model-independent method.

\section{Our improved model-independent method}
\subsection{Our idea}
We found that the contamination from higher-order derivatives remains for large $|f_1|$ and small volume, in mockup data analyses~\cite{Sato:2022qee} assuming the monopole form factor. Our strategy to reduce the contamination is to improve the convergence of the coefficient $f_{m}$ of the form factor $F_{\pi}(q^2)$. Specifically, we change $F_{\pi}(q^2)$ to $S(q^2):=F_{\pi}(q^2)G(q^2)$ by introducing an appropriate function $G(q^2)$ that improves the convergence of $f_{m}$. In Eq.~(\ref{eq:sumpDTF}), inserting an identity $1=G(q^2)/G(q^2)$ yields $C^{(n)}(t)=\sum_{p}\Delta(t,p)T_{n}(p)S(q^2)/G(q^2)$. From the Taylor expansion of $S(q^2)$, $S(q^2)=\sum_{m=0}^{\infty}s_{m}q^{2m}$, it can be written as $C^{(n)}(t)=s_{0}\tilde{\beta}_{0,n}(t)+s_{1}\tilde{\beta}_{1,n}(t)+s_{2}\tilde{\beta}_{2,n}(t)+\cdots=\sum_{m=0}^{\infty}s_{m}\tilde{\beta}_{m,n}(t)$, where $\tilde{\beta}_{m,n}(t)$ is a known function, $\tilde{\beta}_{m,n}(t):=\sum_{p}\Delta(t,p)T_{n}(p)q^{2m}/G(q^2)$. Therefore, the original model-independent method in Eq.~(\ref{eq:R(t)}) is modified to
\begin{eqnarray}
	R(t)=\alpha'_{1}C^{(1)}(t)+\alpha'_{2}C^{(2)}(t)+h'
        =s_{1}+\sum_{m=3}^{\infty}\qty(\sum_{k=1}^{2}\alpha'_{k}\tilde{\beta}_{m,k}(t))s_{m}
	\label{eq:OurR(t)} ,
\end{eqnarray}
where the parameters $\alpha'_{1},\alpha'_{2},h'$ satisfy the equation with $\beta$ replaced by $\tilde{\beta}$ in Eq.~(\ref{eq:defParaAlphaH}). The important point is changing the expansion function from $F_{\pi}(q^2)$ to $S(q^2)$ and choosing an appropriate function $G(q^2)$ with good convergent $s_{m}$. If we can find such a function, we can reduce the contamination from the second term in Eq.~(\ref{eq:OurR(t)}) over Eq.~(\ref{eq:R(t)}) and obtain the charge radius with small systematic error.

\subsection{Appropriate function $G(q^2)$ and convergence of $s_{m}$}
To discuss the appropriate function for $G(q^2)$ in our method, we temporarily assume that the form factor is the monopole form, $F_{\pi}(q^2)=\frac{1}{1+q^2/M_{\rm{pole}}^2}=\sum_{m=0}^{\infty}(-1/M_{\rm{pole}}^2)^{m}q^{2m}$, and investigate the convergence of $s_{m}$. Considering the function $G(q^2)=1+\sum_{m=1}g_{m}q^{2m}$, the relations
\begin{eqnarray}
    s_{0}=1 ,\hspace{15pt}s_{m}=\qty(-\dfrac{1}{M_{\rm{pole}}^2})s_{m-1}+g_{m} \hspace{15pt} (m\geq1)
    \label{eq:sm_relation}
\end{eqnarray}
for the coefficient $s_{m}$ are obtained from $S(q^2)=F_{\pi}(q^2)G(q^2)$. If $G(q^2)=1$ (original method), $s_{m}\sim\order{(-1/M_{\rm{pole}}^2)^{m}}$.

\par

We can apply many different functions $G(q^2)$ to drop the influence of the higher-order coefficients. For later convenience, let us explore the following two examples:
\begin{itemize}
 \item For the quadratic function, we obtain $s_{m}=(-1/M_{\rm{pole}}^2)s_{m-1} \hspace{3pt} (m\geq3)$ from Eq.~(\ref{eq:sm_relation}). If we find the parameters $g_{1},g_{2}$ satisfying the condition $s_{2}=0$, other factors $s_{m\ge 3}$ vanish exactly. On the other hand, $s_{2}$ can be calculated from $R(t)$ similar to Eq.~(\ref{eq:OurR(t)}) with the different choice of the parameters $\alpha'_{1},\alpha'_{2},h'$ satisfying $\alpha'_{1}\tilde{\beta}_{0,1}+\alpha'_{2}\tilde{\beta}_{0,2}+h'=0$, $\alpha'_{1}\tilde{\beta}_{1,1}+\alpha'_{2}\tilde{\beta}_{1,2}=0$, $\alpha'_{1}\tilde{\beta}_{2,1}+\alpha'_{2}\tilde{\beta}_{2,2}=1$. Hence, by varying $g_{1},g_{2}$, we can look for their optimal values such that $s_{2}=0$.
\item For the logarithm function, $G(q^2)=1+g_{1}'\log(1+g_{2}'q^2)=1+\sum_{m=1}^{\infty}(-(-g_{2}')^{m}/m)g_{1}'q^{2m}$, we obtain $s_{m}\sim\order{(-1/M_{\rm{pole}}^2)^{m-2}{g_{2}'}^2}$ for the convergence of $s_{m}$. Again, the parameters $g_{1}', g_{2}'$ are chosen to satisfy $s_{2}=0$. We emphasize that, in this case, the convergence is improved to $(m-2)$-th power compared to the one in the original method, and it can be adjusted by $g_{2}'$.
\end{itemize}

\section{Lattice simulation and results}
\subsection{Simulation parameters}
We use 2+1 flavor gauge configurations generated by the PACS-CS Collaboration~\cite{Yamazaki:2012hi}
with the Iwasaki gauge action at $\beta=1.90$
and the nonperturbative $\order{a}$-improved Wilson quark action at $c_{\rm{SW}}=1.715$.
The ensemble parameters are shown in Table \ref{table:sim_param}.
The correlation functions are computed with the $Z(2)\otimes Z(2)$ random source~\cite{Boyle:2008yd}
and the value of $t_{\rm{sink}}$ in the 3-point function is set to $22$.

\begin{table}[!h]
\begin{center}
\begin{tabular}{cccccccc}\hline\hline
$\beta$ & $L^3\times T$ & $L$[fm] & $a$[fm] & $M_\pi$[GeV] &
$N_{\rm conf}$ & $N_{\rm meas}$ \\
\hline
1.90 & 32$^3$$\times$48 & 2.9 & 0.090 & 0.51 & 80 & 192 \\
\hline\hline
\end{tabular}
\end{center}
\caption{
The bare coupling ($\beta$), lattice size ($L^3\times T$), physical spatial extent ($L$[fm]), pion masses ($m_\pi$) are tabulated. We represent $N_{\rm conf}$ and $N_{\rm meas}$ as the number of configurations and the number of measurements per configuration, respectively.
}
\label{table:sim_param}
\end{table}

\subsection{Simulation results}

\begin{figure}[!th]
     \centering
     \includegraphics[width=90mm,pagebox=cropbox]{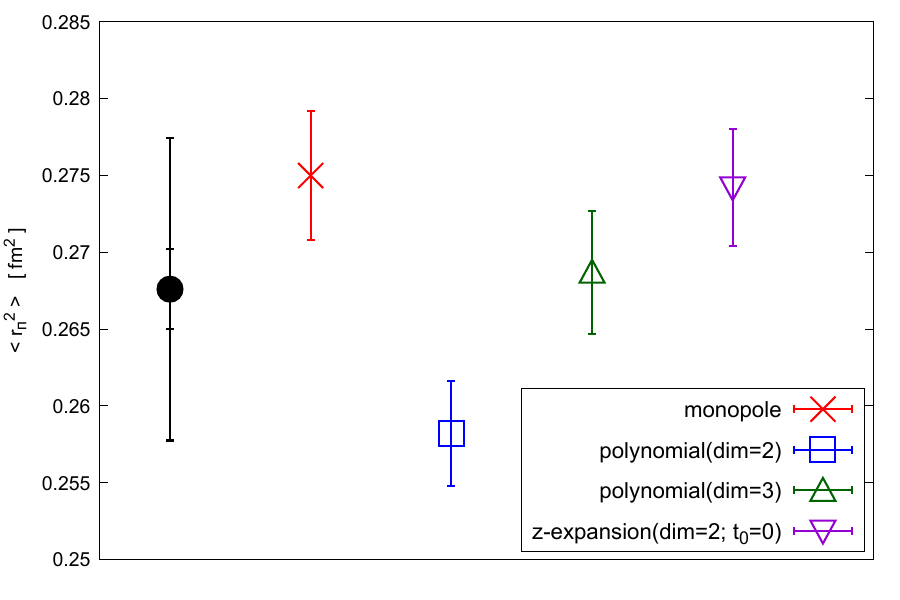}
     \caption{Results for pion charge radius obtained by the traditional method with fitting. The red symbol represents the monopole form result, blue and green represent the quadratic and cubic function results, and purple represents the z-expansion result. The black circle represents the overall results of the traditional method. The inner error is the statistical error and the outer error is the total error.
    \label{fig:Traditional}
    }
\end{figure}

We obtain the pion charge radius from the traditional, original, and our methods. These results are compared using the systematic errors evaluated in each analysis method. 

We use four fitting functions: monopole, quadratic, cubic, and z-expansion, for the traditional method. Each result of the charge radius is shown in Fig.~\ref{fig:Traditional}. It can be seen that each fit function has a different value, which is the systematic error due to the fit ansatz included in the traditional method. In this study, we evaluate the error of the traditional method as follows. The central value and statistical error are determined from the weighted mean and the jackknife error for the four results. The systematic error is the maximum difference between the central value and the value on each form.

\begin{figure}[!th]
    \centering
    \includegraphics[width=75mm,pagebox=cropbox]{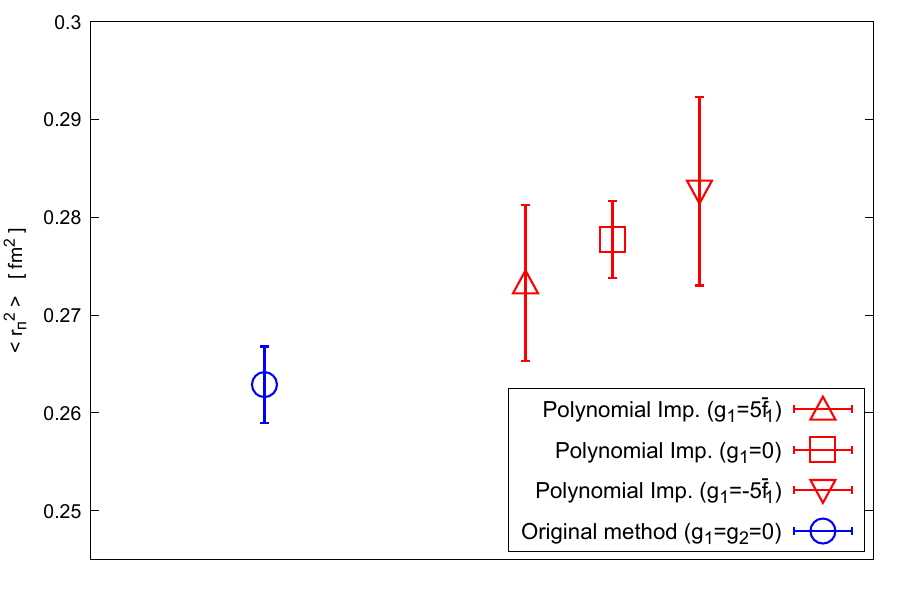}
    \includegraphics[width=75mm,pagebox=cropbox]{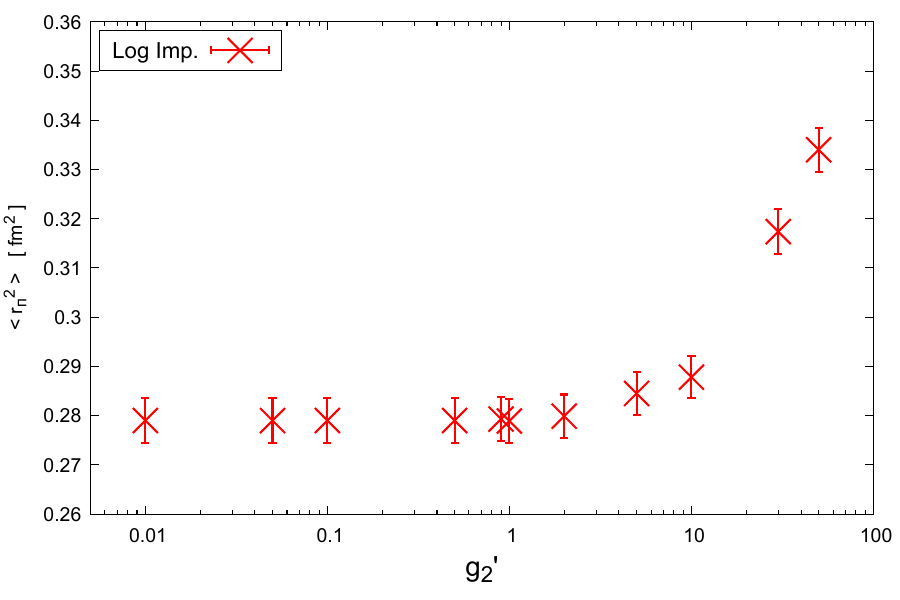}
    \caption{Results for pion charge radius obtained by the model-independent method. Left: Result when $G(q^2)$ is a quadratic function. The blue symbol is the original method result, and the red symbols are our method results at different values of $g_{1}$. Right: Result when $G(q^2)$ is a logarithm function. The horizontal axis is the value of the parameter $g_{2}'$.
    }
  \label{fig:Model-independent}
\end{figure}

Next, we show the results of the model-independent method for the quadratic and logarithmic functions for $G(q^2)$. For the quadratic function $G(q^2)=1+g_{1}q^2+g_{2}q^4$, we fix $g_{1}$ and numerically search $g_{2}$ such that $s_{2}=0$. In this analysis, since the fixed $g_1$ causes a systematic error, the error of $g_{1}$ is evaluated by varying $g_1$ over a sufficiently large range. The results for various values of $g_{1}$ are shown on the left side of Fig.~\ref{fig:Model-independent}. The central value and statistical error are chosen as the values of the result at $g_{1}=0$. The systematic error is estimated by the maximum difference between the central value and results with $g_1$ between $-5\bar{f}_{1}$ and $5\bar{f}_{1}$, where the value of $\bar{f}_{1}$ is determined from the charge radius in the traditional method.

\par

For the logarithm function, $G(q^2)=1+g_{1}'\log(1+g_{2}'q^2)$, similar to the quadratic case, the systematic error coming from the choice of the parameter is evaluated by taking a sufficiently large $g_{2}'$ range with $g_{1}'$ satisfying $s_2 = 0$. The results for various values of $g_{2}'$ are shown on the right side of Fig.~\ref{fig:Model-independent}. When $g_{2}'$ is sufficiently small $g_{2}'\leq1$, the charge radius is constant against $g_{2}'$, and hence, the contribution from higher-order derivatives is considered to be suppressed enough. From this observation, we choose the result at $g_{2}'=1$ as the central value and statistical error for this analysis. The systematic error estimated in the $g_{2}'\leq1$ region is negligible compared to the statistical error.

\par

\begin{figure}[!th]
     \centering
     \includegraphics[width=90mm,pagebox=cropbox]{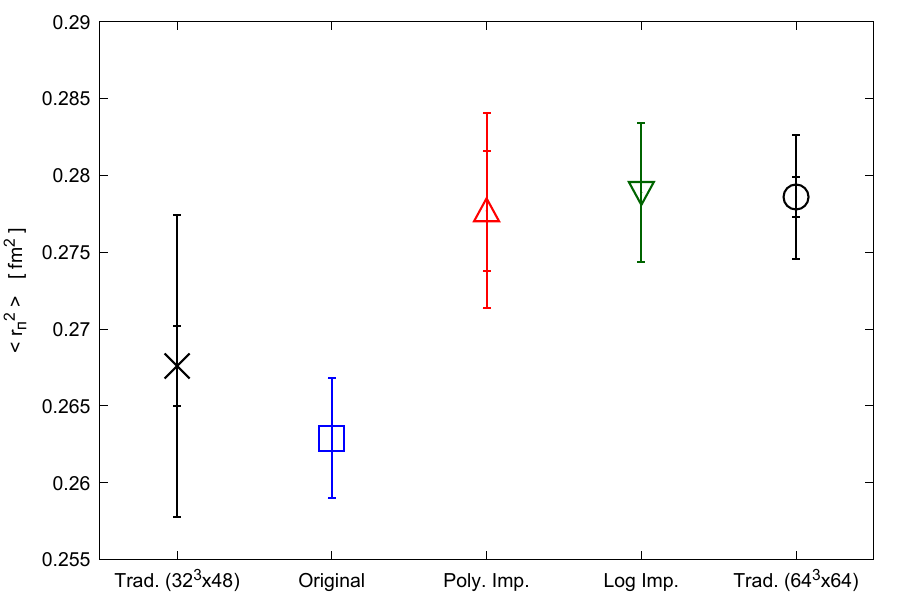}
     \caption{Results of pion charge radius obtained by each analysis method. Black cross and circle are the results of the traditional method for small and large volumes; colored symbols are the results of the model-independent method.
    \label{fig:result}
    }
\end{figure}

The above results are summarized in Fig.~\ref{fig:result}. These results show that the model-independent method's error is smaller than the traditional method’s error on this volume. We also find that there is a difference between the results for the original method and our method. This implies that our method can suppress the finite volume effect, because our method is consistent with the result of the traditional analysis on a larger volume of the $64^3\times64$ lattice\footnote{Although the details are not shown in this proceedings, we also performed a similar analysis on the large volume ($64^3\times64$) configuration, which will be presented in our forthcoming paper.}.

\section{Summary}
We discuss the improvement of the original model-independent method to obtain the pion charge radius in lattice QCD. We propose a new method that reduces the high-order contribution included in the original model-independent method. We also apply the method to actual lattice QCD data at $m_{\pi}=0.51$ GeV and find that the error is reduced from the traditional method and it can suppress the finite volume effect.

\section*{Acknowledgments}
Numerical calculations in this work were performed on Oakforest-PACS and Wisteria/BDEC-01 (Odyssey) in Joint Center for Advanced High Performance Computing, and on Cygnus in Center for Computational Sciences at the University of Tsukuba under Multidisciplinary Cooperative Research Program of Center for Computational Sciences, University of Tsukuba. The calculation employed OpenQCD system\footnote{http://luscher.web.cern.ch/luscher/openQCD/}. This work was supported in part by Grants-in-Aid for Scientific Research from the Ministry of Education, Culture, Sports, Science and Technology (No. 19H01892, 23H01195), MEXT as “Program for Promoting Researches on the Supercomputer Fugaku” (JPMXP1020230409), and JST, The Establishment of University Fellowships towards the creation of Science Technology Innovation, Grant Number JPMJFS2106. This work was supported by the JLDG constructed over the SINET5 of NII. 

\bibliography{reference}

\providecommand{\href}[2]{#2}\begingroup\raggedright\begin{thebibliography}{1}

\bibitem{Aglietti:1994nx}
U.~Aglietti, G.~Martinelli and C.~T. Sachrajda, \href{https://doi.org/10.1016/0370-2693(94)00053-0}{\emph{Phys. Lett. B} {\bfseries 324} (1994) 85} [\href{https://arxiv.org/abs/hep-lat/9401004}{{\ttfamily hep-lat/9401004}}].

\bibitem{Lellouch:1994zu}
{\scshape UKQCD} collaboration, \href{https://doi.org/10.1016/0550-3213(95)00180-Z}{\emph{Nucl. Phys. B} {\bfseries 444} (1995) 401} [\href{https://arxiv.org/abs/hep-lat/9410013}{{\ttfamily hep-lat/9410013}}].

\bibitem{Bouchard:2016gmc}
C.~Bouchard, C.~C. Chang, K.~Orginos and D.~Richards, \href{https://doi.org/10.22323/1.256.0170}{\emph{PoS} {\bfseries LATTICE2016} (2016) 170} [\href{https://arxiv.org/abs/1610.02354}{{\ttfamily 1610.02354}}].

\bibitem{Feng:2019geu}
X.~Feng, Y.~Fu and L.-C. Jin, \href{https://doi.org/10.1103/PhysRevD.101.051502}{\emph{Phys. Rev. D} {\bfseries 101} (2020) 051502} [\href{https://arxiv.org/abs/1911.04064}{{\ttfamily 1911.04064}}].

\bibitem{Sato:2022qee}
K.~Sato, H.~Watanabe and T.~Yamazaki, \href{https://doi.org/10.22323/1.430.0122}{\emph{PoS} {\bfseries LATTICE2022} (2023) 122} [\href{https://arxiv.org/abs/2212.00207}{{\ttfamily 2212.00207}}].

\bibitem{Yamazaki:2012hi}
T.~Yamazaki, K.-i. Ishikawa, Y.~Kuramashi and A.~Ukawa, \href{https://doi.org/10.1103/PhysRevD.86.074514}{\emph{Phys. Rev. D} {\bfseries 86} (2012) 074514} [\href{https://arxiv.org/abs/1207.4277}{{\ttfamily 1207.4277}}].

\bibitem{Boyle:2008yd}
{\scshape RBC-UKQCD} collaboration, \href{https://doi.org/10.1088/1126-6708/2008/07/112}{\emph{JHEP} {\bfseries 07} (2008) 112} [\href{https://arxiv.org/abs/0804.3971}{{\ttfamily 0804.3971}}].

\end{thebibliography}\endgroup

\end{document}